\documentclass{aa}
\usepackage{graphicx}
\begin{document}
   \title{ISO spectroscopy of star formation and active nuclei in the 
   luminous infrared galaxy \object{NGC\,6240}\thanks
   {Based on observations with ISO, an ESA project with instruments funded
   by ESA member states (especially the PI countries: France, Germany, the
   Netherlands, and the United Kingdom) with the participation of ISAS and
   NASA.}
   }

   \subtitle{}

   \author{D. Lutz \inst{1}
          \and
          E. Sturm \inst{1}
          \and
          R. Genzel \inst{1}
          \and
          H.W.W. Spoon \inst{2}
          \and
          A.F.M. Moorwood \inst{3}
          \and
          H. Netzer \inst{4}
          \and
          A. Sternberg \inst{4}
          }

   \offprints{D. Lutz}

   \institute{Max-Planck-Institut f\"ur extraterrestrische Physik, 
              Postfach 1312, 85741 Garching, Germany\\
              \email{lutz@mpe.mpg.de, sturm@mpe.mpg.de, genzel@mpe.mpg.de}
         \and
             Kapteyn Institute, P.O. Box 800, 9700 AV Groningen, the
             Netherlands\\
             \email{spoon@astro.rug.nl}
         \and
             European Southern Observatory, Karl-Schwarzschild-Str. 2,
             85748 Garching, Germany\\
             \email{amoor@eso.org}
         \and
             School of Physics and Astronomy and Wise Observatory, 
             Raymond and Beverly Sackler Faculty of Exact Sciences,
             Tel Aviv University, Ramat Aviv,
             Tel Aviv 69978, Israel\\
             \email{netzer@wise1.tau.ac.il, amiel@wise1.tau.ac.il}
             }

   \date{Received 23 March 2003; accepted 22 July 2003}

   \abstract{We present Infrared Space Observatory mid- and far-infrared 
   spectroscopy of the merging galaxy \object{NGC\,6240}, an object presenting 
   many aspects of importance for the role of star formation and AGN activity 
   in [ultra]luminous infrared galaxies. The mid-infrared spectrum shows 
   starburst
   indicators in the form of low excitation fine-structure line emission and 
   aromatic `PAH' features. A strong high excitation [O\,IV] line
   is observed which most likely originates in the Narrow Line Region of an
   optically obscured AGN. \object{NGC\,6240} shows extremely powerful
   emission in the pure rotational lines of molecular hydrogen.
   We argue that this emission is mainly due to shocks in its 
   turbulent central gas component and its starburst superwind. The total
   shock cooling in infrared emission lines accounts for $\sim$0.6\% of the
   bolometric luminosity, mainly through rotational H$_2$ emission and the
   [O\,I]\,63$\mu$m line. We analyse several ways of
   estimating the luminosities of the starburst and the AGN in NGC\,6240
   and suggest that the contributions to its bolometric luminosity are
   most likely in the range 50-75\% starburst and 25-50\% AGN.
   }

\titlerunning{Starburst and AGN in the infrared spectrum of 
              \object{NGC\,6240}}
\authorrunning{D. Lutz et al.}

   \maketitle

\section{Introduction}
The double nucleus merging galaxy \object{NGC\,6240} is a relatively nearby
(cz=7339\,km/s, d=98\,Mpc) infrared-luminous system. 
At $L_{\rm IR}=6.05\times 10^{11}L_\odot$  (see e.g. Sanders \& Mirabel 
\cite{sanders96} for the definition of the 8-1000$\mu$m luminosity 
$L_{\rm IR}$), 
it is almost an ultraluminous infrared galaxy and thus allows 
detailed studies of many aspects of the phenomenon of (ultra)luminous 
infrared galaxies ([U]LIRGs). For the dusty system \object{NGC\,6240}, 
the infrared luminosity is identical to the bolometric
luminosity to good approximation, the optical/near-infrared and
X-ray regions observed to be minor contributors. In the infrared 
selected sample studied spectroscopically with the Infrared Space 
Observatory ISO by Genzel et al. (\cite{genzel98}) it is 
among the sources with indications for an active galactic nucleus (AGN) in 
the mid-infrared emission line spectrum, similar indications can be
obtained from the low resolution mid-infrared diagnostic diagram of 
Laurent et al. (\cite{laurent00}). Several X-ray
studies have firmly established the presence of powerful AGN activity
(e.g., Iwasawa \& Comastri \cite{iwasawa98}; Vignati et al. \cite{vignati99}),
in fact occuring in both nuclei (Komossa et al. \cite{komossa03}). Coexistence
of star formation and AGN activity is a feature of many [ultra]luminous
infrared galaxies. The proximity of \object{NGC\,6240} and the extensive
set of multi-wavelength observations make it a unique target for
studying the quantitative contributions of the AGN and star formation activity
in a [U]LIRG
with a relatively powerful AGN. Such studies are not only of interest for an
understanding of the local [U]LIRG population, but also for the high
redshift infrared and submm selected populations to which the [U]LIRGs
likely form the closest available local analogs.  

\begin{table*}
\begin{tabular}{llllcr}\hline
Obs\_ID &Date       &AOT&Proposal          &$N_{Ranges}$&Duration\\ 
        &UT         &   &                  &            &h:mm:ss\\ \hline
09700106&22-FEB-1996&S02&RGENZEL\_MPEXGAL2 &16          &5:06:22\\
09700207&22-FEB-1996&S02&RGENZEL\_MPEXGAL2 &11          &2:20:14\\
31101401&23-SEP-1996&S02&ASTERNBE\_FEGALAXI&2           &0:56:40\\
31801036&29-SEP-1996&S02&RGENZEL\_MPEXGAL3 &7           &2:53:52\\
48101201&11-MAR-1997&S02&PVDWERF\_SHOCKS   &7           &1:49:02\\
83300305&25-FEB-1998&S02&EEGAMI\_H2        &3           &1:15:26\\
83500308&27-FEB-1998&S02&EEGAMI\_H2        &3           &1:00:34\\
84400107&08-MAR-1998&S02&EEGAMI\_H2        &8           &2:06:38\\
86600118&30-MAR-1998&S02&DLEVINE\_FILLFUP  &3           &0:42:02\\
27801108&21-AUG-1996&L01&HSMITH\_IRBGALS   &n.a.        &1:09:04\\
81000771&02-FEB-1998&L02&HSMITH\_IRBGALS   &n.a.        &0:20:38\\
81000872&02-FEB-1998&L02&HSMITH\_IRBGALS   &n.a.        &0:20:38\\
81000973&02-FEB-1998&L02&HSMITH\_IRBGALS   &n.a.        &0:20:38\\
81001074&02-FEB-1998&L02&HSMITH\_IRBGALS   &n.a.        &0:20:38\\
81600475&08-FEB-1998&L02&HSMITH\_IRBGALS   &n.a.        &0:20:40\\
81600676&08-FEB-1998&L02&HSMITH\_IRBGALS   &n.a.        &0:20:40\\
\hline
\end{tabular}
\caption{Log of the ISO observations of \object{NGC\,6240}
presented in this paper. The number
of wavelength ranges includes `serendipitous' data obtained by a different
detector in different order simultaneously with a requested range, but 
excludes data outside the nominal SWS wavelength bands.}
\label{tab:obs}
\end{table*}

A second important question relates to the evolution of merging galaxies
like \object{NGC\,6240}. Recent near-infrared dynamical studies (Genzel et al.
\cite{genzel01}, Tacconi et al. \cite{tacconi02}) suggest that
merging ultraluminous infrared galaxies as a class will evolve into
moderate mass, disky-type elliptical galaxies. \object{NGC\,6240} stands
out in this group by its very high stellar velocity dispersion
(Lester \& Gaffney \cite{lester94}, Doyon et al. \cite{doyon94})
suggesting a very massive object. Spatially resolved dynamical studies using
CO (cold gas; Tacconi et al. \cite{tacconi99}), H$_2$ (hot gas; van der Werf
et al. \cite{vdwerf93}; Tecza et al. \cite{tecza00}), and stellar
absorption features (Tecza et al. \cite{tecza00}) reveal complex, not
fully relaxed dynamics with maximum velocity dispersion between the two
nuclei. The extraordinary luminosity of \object{NGC\,6240} in the
near-infrared rovibrational lines of molecular hydrogen (Joseph et al.
\cite{joseph84}) may be partly linked to this complex dynamics, as well 
as to shocked emission
in the superwind flow that is driven by the \object{NGC\,6240} starburst 
(van der Werf et al. \cite{vdwerf93}).

This paper presents ISO mid- and far-infrared spectroscopic  observations of 
\object{NGC\,6240} and discusses implications for its starburst and AGN
activity, and the nature of the powerful H$_2$ emission.

\section{Observations and Reduction}

\begin{figure*}
\centering
\includegraphics[width=0.975\textwidth]{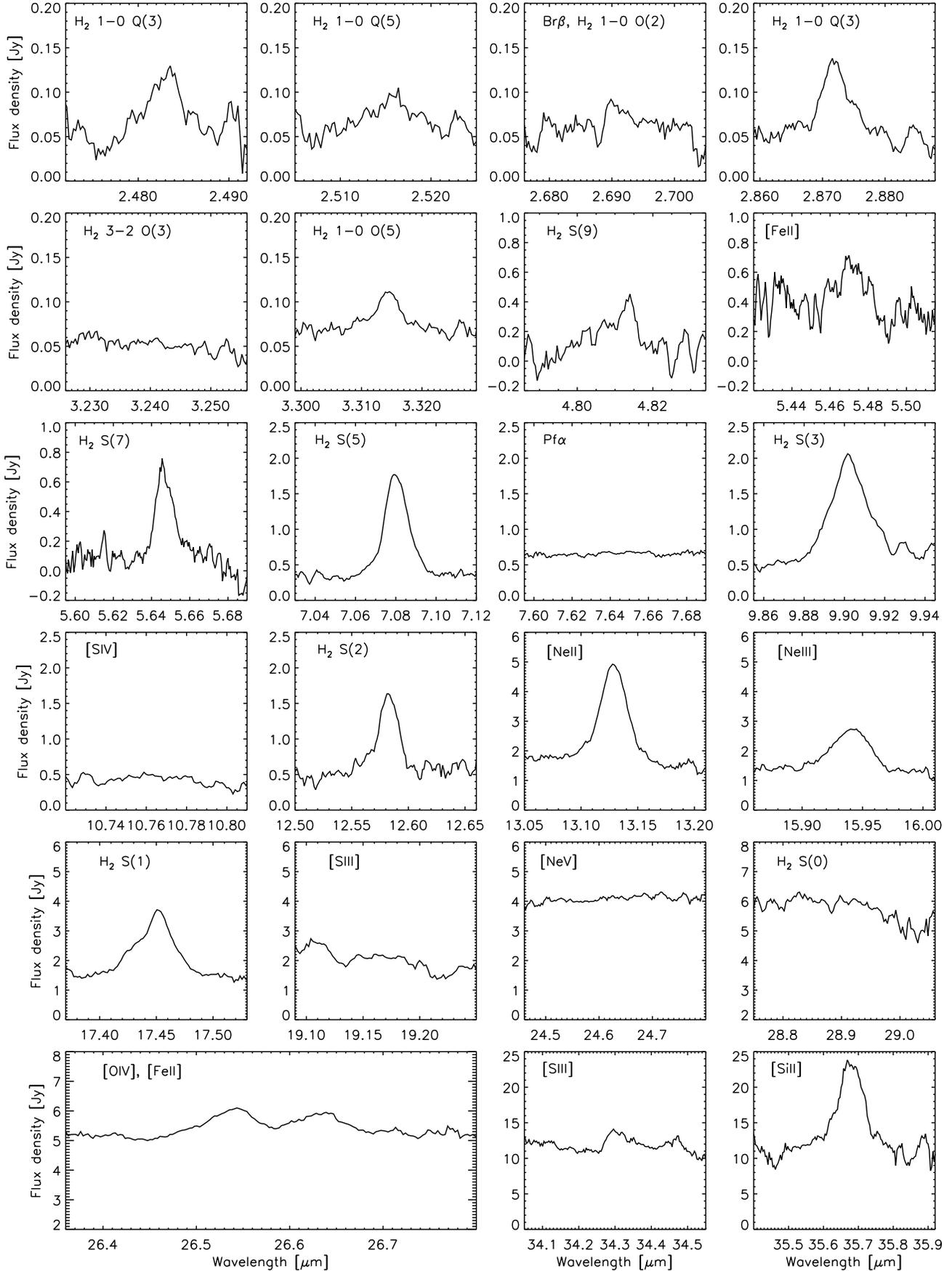}
\caption{ISO-SWS spectra of the lines detected in \object{NGC\,6240}. Some 
of the ranges from which important upper limits are derived are also shown.}
\label{fig:swsspec}
\end{figure*}

We have retrieved from the ISO archive all observations of 
\object{NGC\,6240} that were obtained with the ISO Short Wavelength
Spectrometer SWS (de Graauw et al. \cite{degraauw96}). Table~\ref{tab:obs}
lists the basic data. All observations were obtained in the SWS02 mode
which produces short spectral segments at the nominal SWS resolving power
($\sim$2000), centered on selected wavelengths.
The data were reduced in the OSIA\footnote{OSIA is a joint development 
of the SWS consortium. Contributing institutes are SRON, MPE, KUL and the 
ESA Astrophysics Division.} 3.0 software framework which adopts the SWS
calibration files of the offline processing software version 10.1. Standard 
reduction procedures were 
used, putting particular emphasis on identifiying and eliminating individual
noisy detectors and scans affected by signal jumps. Repeated observations of
the same range were offset by small values to a consistent median
flux. Single valued spectra were then produced by kappa-sigma clipping 
and averaging the many independent measurements inside a resolution
element. A resolution element of $\lambda$/2000 was used for this step. 
While leading to a slight degradation of the nominal SWS resolution, 
this is favourable for signal-to-noise reasons and does not affect the wide
line profiles of \object{NGC\,6240} noticeably. Figure~\ref{fig:swsspec}
shows the segments containing the detected lines and some ranges providing
important upper limits. The
spectra were rebinned with R$\sim$2000 and are five times oversampled with
respect to that resolution. Table~\ref{tab:fluxes} lists the derived fluxes
(estimated to be accurate to 20-30\%),
limits, and line parameters.

All observations were centered on the region of the \object{NGC\,6240}
double nucleus (Fried \& Schulz \cite{fried83}). The SWS apertures 
(between $14\arcsec\times 20\arcsec$ and
$20\arcsec\times 33\arcsec$) include both nuclei as well as the extended
region of emission in the near-infrared rovibrational lines of molecular
hydrogen (e.g. van der Werf et al. \cite{vdwerf93}). Partial analyses of
the SWS spectra of \object{NGC\,6240} using
only a subset of the data with earlier calibrations and addressing only 
certain aspects are presented
in Genzel et al. (\cite{genzel98}) and Egami (\cite{egami98};
\cite{egami99}).

\begin{figure*}
\centering
\includegraphics[width=\textwidth]{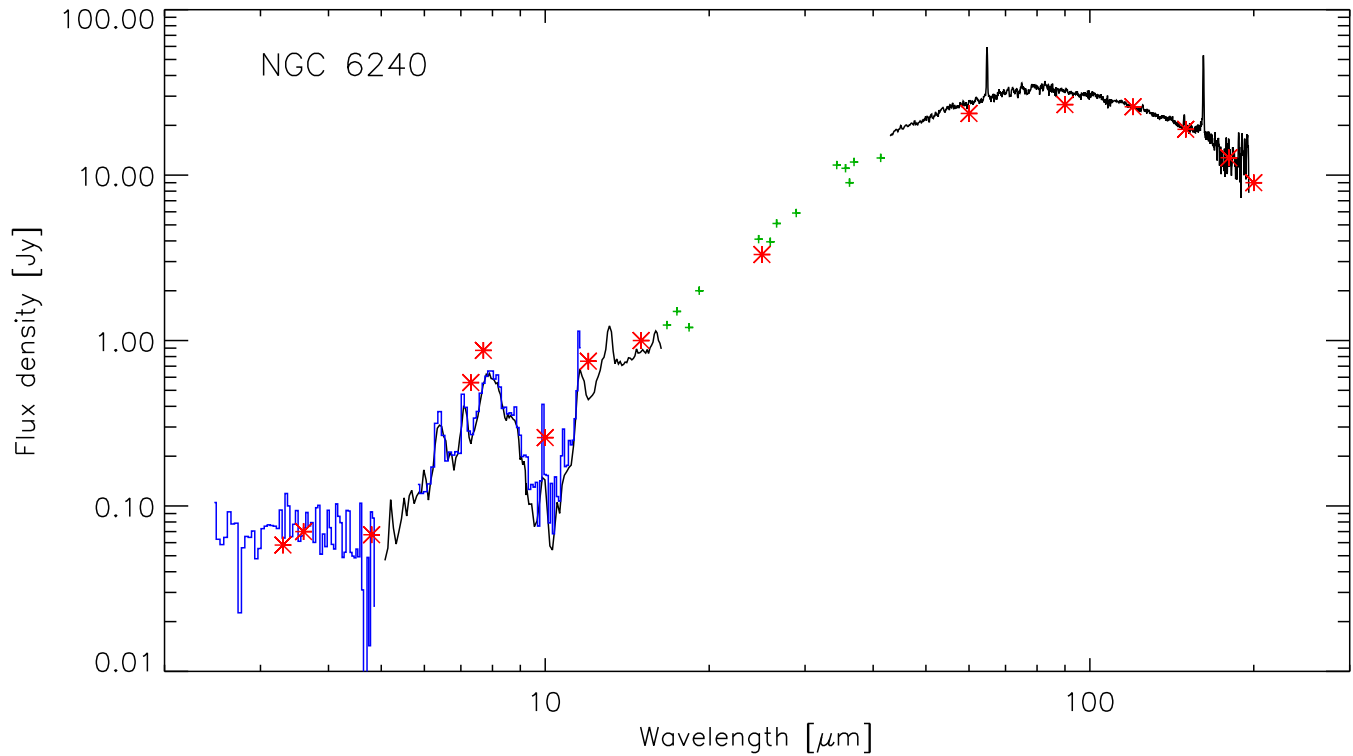}
\caption{Infrared spectral energy distribution of \object{NGC\,6240} from
ISO observations. Continuous lines show the ISOCAM-CVF and LWS spectra,
histograms the ISOPHOT-S spectrum. Small crosses indicate continua
from the SWS spectra at $>16\mu$m, where the relative zodiacal light 
contribution is small. Asterisks indicate the
broad band photometry of Klaas et al. (\cite{klaas97},\cite{klaas01})}
\label{fig:sed}
\end{figure*}

We have also reduced an ISO-LWS 43--196$\mu$m full spectrum (L01 mode)
as well as 6 grating line scan observations (L02 mode) of the 
[N\,III] 57$\mu$m
line (see Table~\ref{tab:obs}). The data were first reduced in 
LIA\footnote{LIA is a joint development of the ISO-LWS Instrument Team at
Rutherford Appleton Laboratories (RAL, UK - the PI Institute) and the Infrared
Processing and Analysis Center (IPAC/Caltech, USA).}
version 8.1, based on offline processing software version 10.1 calibration 
files. In particular dark current
subtraction and absolute responsivity correction were done interactively.
Following the LWS recipes for faint sources we have further reduced the data
using ISAP\footnote{The ISO Spectral Analysis Package (ISAP) is a 
joint development by 
the LWS and SWS Instrument Teams and Data Centers. Contributing institutes 
are CESR, IAS, IPAC, MPE, RAL and SRON.} 
version 2.1. We deleted glitches and memory tails, and 
averaged multiple scans using ISAP's standard clip \& mean option with bin 
sizes preserving the sample spacing. 
To produce the continuous full grating spectrum shown in Fig.~\ref{fig:sed}
we have shifted the 10 LWS01 sub-spectra in offset mode by values 
between 5 and 20 percent (30\% for detector LW5), using detector LW04 
($\approx$ 140 to 170 $\mu$m) as reference (``good'') detector. 
With an average effective LWS beam size of 77$\arcsec$ (Gry et al. 
\cite{gry02})
all relevant regions of NGC\,6240 are covered. Fluxes or upper limits are
included in Table~\ref{tab:fluxes}. All measured lines are unresolved at 
the resolution of LWS grating spectra (150 to 300, $\approx$ 1500 km/s).

We have reanalysed the ISOPHOT-S spectrum already presented in
Genzel et al. (\cite{genzel98}) and Rigopoulou et al. (\cite{rigo99})
using PIA\footnote{PIA is a joint development by the ESA Astrophysics division
and the ISO-PHT consortium} version 9.0.1. The spectrum was obtained on 1996 
February 15 using rectangular chopped mode (chopper throw of 
180$\arcsec$). The pure on-source integration time was 512\,s.
Steps in the data reduction included: 1) 
deglitching on ramp level. 2) subdivision of ramps in four sections 
of 32 non destructive read-outs. 3) ramp fitting to derive 
signals. 4) masking of bad signals by eye-inspection. 5) kappa sigma 
and min/max clipping on remaining signal distribution. 6)
determination of average signal per chopper plateau. 7) masking or 
correction of bad plateaux by eye-inspection. 8) background
subtraction using all but the first four plateaux. 9) finally, flux
calibration, using the signal dependent spectral response function.
The absolute calibration is accurate to within 20\%.
Figure~\ref{fig:sed} shows the 
infrared spectral energy distribution of \object{NGC\,6240} combining the 
ISOPHOT-S spectrum, the ISOCAM-CVF data of Laurent et al. (\cite{laurent00} and
priv. communication), 
the LWS data and published broadband photometry.

\begin{table*}
\begin{tabular}{lrrrr}\hline
Transition&$\lambda_{\rm Rest}$&$\lambda_{\rm Obs}$&Flux           &FWHM\\
          &$\mu$m          &$\mu$m&10$^{-20}$\,W\,cm$^{-2}$&km/s\\ \hline
H$_2$ 1-0 Q(3) & 2.4237     &2.4832         &1.55           &540\\ 
H$_2$ 1-0 Q(5) & 2.4548     &2.5159         &0.83           &604\\
H Br$\beta$    & 2.6259     &               &$<$0.90        &   \\
H$_2$ 1-0 O(3) & 2.8025     &2.8720         &1.59           &548\\
H$_2$ 2-1 O(3) & 2.9741     &               &$<$0.35        &   \\
H$_2$ 3-2 O(3) & 3.1638     &               &$<$0.30        &   \\
H$_2$ 1-0 O(5) & 3.2350     &3.3144         &0.56           &436\\
H Br$\alpha$   & 4.0523     &               &$<$2.5         &   \\
H$_2$ 0-0 S(11)& 4.1813     &               &$<$2.6         &   \\
H$_3^+1,2,3_{+1}\rightarrow 3,3$& 4.350\ &  &$<$1.9         &   \\
H$_2$ 0-0 S(9) & 4.6946     &4.8114         &3.90           &600$^1$\\
{[Fe\,II]}     & 5.3402     &5.4707         &5.46           &875\\
H$_2$ 0-0 S(7) & 5.5112     &5.6469         &7.27           &636\\
H$_2$ 0-0 S(5) & 6.9095     &7.0803         &12.20          &580\\
H Pf$\alpha$   & 7.4599     &               &$<$0.6         &   \\
H$_2$ 0-0 S(3) & 9.6649     &9.9021         &9.24           &641\\
{[S\,IV]}      &10.5105     &               &$<$1.1         &   \\
H$_2$ 0-0 S(2) &12.2786     &12.5821        &4.35           &484\\
{[Ne\,II]}     &12.8136     &13.1288        &16.20          &630\\
{[Ne\,V]}      &14.3217     &               &$<$0.8         &   \\
{[Ne\,III]}    &15.5551     &15.9405        &6.31           &675\\
H$_3^+5,0\rightarrow 4,3$&16.331\ &         &$<$1.0         &   \\ 
H$_2$ 0-0 S(1) &17.0348     &17.4497        &7.83           &657\\
{[Fe\,II]}     &17.9360     &               &$<$1.6         &   \\
{[S\,III]}     &18.7130     &               &$<$2.0         &   \\
{[Ne\,V]}      &24.3175     &               &$<$0.6         &   \\
{[S\,I]}       &25.2490     &               &$<$1.2         &   \\
{[O\,IV]}      &25.8903     &26.5384        &2.33           &672\\            
{[Fe\,II]}     &25.9883     &26.6322        &1.85           &653\\
H$_2$ 0-0 S(0) &28.2188     &               &$<$2.2         &   \\
{[S\,III]}     &33.4810     &34.3091        &3.57           &503\\
OH 1/2-3/2     &34.6164     &               &$\tau <0.5$    &no absorption\\
{[Si\,II]}     &34.8152     &35.6786        &26.0           &731\\
{[Fe\,II]}     &35.3487     &               &$<$2.4         &   \\
{[Ne\,III]}    &36.0135     &               &$<$2.6         &   \\
H$_2$O $6_{43}\rightarrow 5_{32}$&40.3367&  &$<$3.7         &   \\
{[O\,III]}     &51.8145     &               &$<$14.0        &   \\
{[N\,III]}     &57.3170     &               &$<$4.0         &   \\
{[O\,I]}       &63.1837     &64.744         &72.00          &unresolved\\
{[O\,III]}     &88.3560     &               &$<$6.0         &   \\
{[N\,II]}      &121.8976     &124.95         &2.3           &unresolved\\
H$_2$O $3_{13}\rightarrow 2_{02}$&138.5272& &$<$1.7         &   \\
{[O\,I]}       &145.5254    &148.99         &4.00           &unresolved\\
CO 17-16       &153.2667    &               &$<$1.8         &   \\
{[C\,II]}      &157.7409    &161.598        &29.00          &unresolved\\
\hline
\end{tabular}
\caption{Line fluxes and limits measured with SWS and LWS for NGC\,6240. 
Upper limits for line fluxes are for an adopted FWHM of 600km/s (SWS) which
corresponds to an unresolved line for LWS.
{$^1$} FWHM enforced for the gaussian fit used to derive the line flux.} 
\label{tab:fluxes}
\end{table*}

\section{Results}
The mid-infrared fine-structure emission line spectrum of \object{NGC\,6240} 
resembles the spectra of
starbursting infrared galaxies (Thornley et al. \cite{thornley00};
F\"orster Schreiber et al. \cite{foerster01}; 
Verma et al. \cite{verma03}) in showing strong [Ne\,II] 12.81$\mu$m
emission and a moderate excitation as probed by the ratio 
[Ne\,III] 15.55$\mu$m / [Ne\,II] 12.81$\mu$m. It differs from a typical
starburst spectrum in the enormous strength of the
rotational emission lines of warm molecular hydrogen, and in the fairly
strong emission in [O\,IV] 25.89$\mu$m, a line which is one of the strongest 
tracers of the mid-infrared spectra of AGN (Sturm et al. \cite{sturm02}),
but very weak or absent in the spectra of starbursts (Lutz et al. 
\cite{lutz98}; Verma et al. \cite{verma03}).

The overall spectral energy distribution and mid-infrared low resolution 
spectrum (Fig.~\ref{fig:sed}) also resemble star forming galaxies, with
prominent emission in the mid-infrared aromatic `PAH' features. The rotational
lines of molecular hydrogen are so strong that some of them are easily
detected in the low resolution spectra. 

\subsection{Detection of AGN Narrow Line Region emission}
\label{sect:nlrdet}

\begin{figure}
\includegraphics[width=\columnwidth]{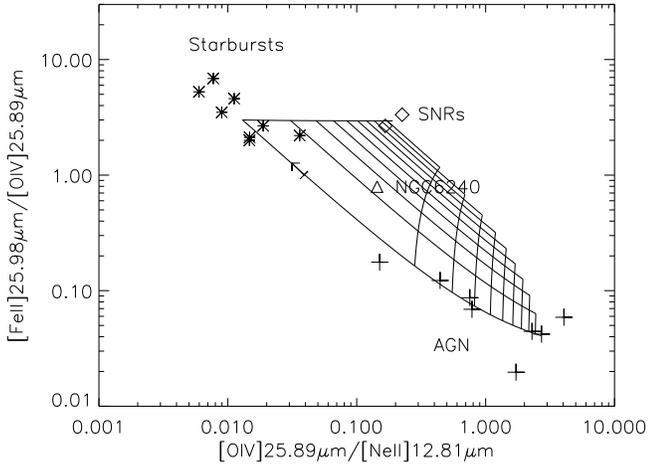}
\caption{Fine-structure emission line ratios for \object{NGC\,6240}, compared
to a number of bright AGN from Sturm et al. (\cite{sturm02}), the starbursts of
Verma et al. (\cite{verma03}, excluding low metallicity dwarfs),
and the supernova remnants \object{IC\,443} and  \object{RCW\,103} 
(Oliva et al. \cite{oliva99a}, \cite{oliva99b}). The overplotted grid 
represents simple mixing by a linear superposition of 
starburst, SNR, and AGN line spectra. The loci of the grid lines 
indicate where the contribution to [Ne\,II] is 0\%, 10\%, 20\%, \ldots\ 
for supernova remnants and for AGN. The percentage contribution to other lines
will be different. Note that the four leftmost AGN 
themselves have evidence (PAH emission) for significant star formation 
inside the SWS beam.}
\label{fig:fsratios}
\end{figure}

The detection of [O\,IV] at $\sim$15\% of the [Ne\,II] flux is clearly
above the regime of starbursts and approaches values for AGN
(Fig.~\ref{fig:fsratios}). Under the assumption of simple mixing of AGN
and starburst emission, the AGN will then contribute little to [Ne\,II] but 
will dominate [O\,IV]. A word of caution is 
in place here about possible other origins of [O\,IV]. The very low level
[O\,IV] emission seen in starbursts is not generally attributed to AGN but 
to either
supernova or wind-related ionizing shocks or very hot stars 
(Lutz et al. \cite{lutz98}, Schaerer \& Stasinska \cite{schaerer99}). 
Given the dynamics and the superwind activity of \object{NGC\,6240}
and the extremely strong and likely shock-related molecular hydrogen
emission, it is conceivable that, compared to a normal
starburst, ionizing shocks producing [O\,IV] are at a significantly elevated 
level as well. One way to probe for such a contribution
is to look for other emissions that are strong in ionizing shocks. An
obvious choice here is [Fe\,II] which is amply emitted in
partially ionized zones of shocks (e.g. Graham et al. \cite{graham87};
Hollenbach \& McKee \cite{hollenbach89}) and additionally boosted by shock 
destruction of grains (e.g. Jones et al. \cite{jones96}; Oliva et al.
\cite{oliva99a}, \cite{oliva99b}) onto which 
Fe is normally depleted. Indeed both starbursts and
supernova shocks observed with ISO show strong emission in the 
[Fe\,II] 25.98$\mu$m line that is neighbouring [O\,IV]. Similar ratios 
of the two lines are observed in starbursts and SNRs 
(Fig.~\ref{fig:fsratios}). In AGN, however,
the [Fe\,II] emission from shocked or UV/X-ray irradiated partially 
ionized zones is (relatively) fainter
compared to the strong [O\,IV] emission from the NLR. The [Fe\,II] emission
in \object{NGC\,6240} is fainter relative to [O\,IV] than in
starbursts or supernova remnants which suggests the AGN to dominate
[O\,IV]. This is quantified in the simple mixing model shown in 
Fig.~\ref{fig:fsratios} which assumes the spectrum to be a linear
superposition of a starburst, an AGN, and a SNR spectrum. At the location
of \object{NGC\,6240}, the contributions to [Ne\,II] are then about 82\%
starburst, 14\% SNR and 4\% AGN. Because of the strongly different intrinsic
[O\,IV]/[Ne\,II] ratios, the contributions to [O\,IV] are very different:
7\% starburst, 18\% SNR, and a dominant 75\% AGN.
This nominal mix as well as the complexity of the excitation 
mechanisms for these lines clearly allow some of the strong [O\,IV] emission
to be due to a high level of non-AGN shock activity, but we will proceed 
in the following with the assumption that [O\,IV] is mainly tracing the AGN
proper. 

The higher ionization potential NLR lines or coronal lines are weak
or absent in  \object{NGC\,6240}. No near-infrared coronal line detections 
have been reported in the literature. This could be due to extinction.
A limit of $<$0.26 can be derived for the extinction-insensitive ratio of 
[Ne\,V] 24.3$\mu$m and the neighbouring [O\,IV] 25.89$\mu$m line. This 
places the NLR of \object{NGC\,6240} towards the 
low end but within the excitation range of the Seyferts studied by Sturm et al.
(\cite{sturm02}), who measured ratios in the range from 0.18 to 0.37. The data
are consistent with [O\,IV] being AGN dominated. 

Egami (\cite{egami99}) suggested an overall dominance of the AGN contribution
to the \object{NGC\,6240} fine-structure line emission, not only for
[O\,IV] but also for the low excitation lines like [Ne\,II]. To make this 
consistent with typical AGN NLR line ratios (e.g., Sturm et al. 
\cite{sturm02}), he invoked a model where 
the ionization structure of the NLR follows an onion-like layered pattern from
high excitation inside to low excitation outside, with obscuration
selectively blocking the higher excitation inner parts. High obscuration 
specifically towards part of the NLR is conceivable, also given the large 
amounts of gas in the central region of \object{NGC\,6240}. We consider an 
extreme effect of obscuration on the line ratios unlikely, however, because 
of the non-ideal structure of real NLRs (e.g. Capetti et al. \cite{capetti97})
which does not show perfect layering of ionisation 
stages as in idealized spherically symmetric photoionisation models. 
In addition, such a scenario
would either require the same peculiar geometry to apply for both AGN,
or insignificance of one of them, which is in contrast to their modest 
hard X-ray flux ratio of $\sim$3 (Komossa et al. \cite{komossa03}).

The combined evidence favours a model where the lower ionisation lines are
dominated by starburst emission but [O\,IV] by the NLR(s) of the AGN(s).  
The absolute 
flux of [O\,IV] from the NLR may still be affected by obscuration. NLR
obscuration
cannot be constrained from the data that are lacking extinction-sensitive
ratios which are purely tracing the NLR. 
In principle, the [O\,III] nondetections (Table~\ref{tab:fluxes}) also 
might constrain properties of the obscured Narrow Line Region 
(e.g. [O\,III] 88$\mu$m / [O\,IV] 26$\mu$m
$<$ 2.6). However, our ignorance about density and ionization parameter of this
region prevents converting this into an obscuration constraint.

\subsection{Properties of the Starburst}
\label{sect:sbprop}

At an excitation ratio [Ne\,III] 15.55$\mu$m / [Ne\,II] 12.81$\mu$m of 0.39, 
\object{NGC\,6240} is well within the range covered by normal and high
metallicity starburst galaxies (Thornley et al. \cite{thornley00}; Verma et al.
\cite{verma03}). This is still true when lowering the ratio by correcting
for the likely noticeable contamination of [Ne\,III] by AGN emission. 
Estimated from AGN [Ne\,III]/[O\,IV] ratios (Sturm et al. 
\cite{sturm02}) and the observed [O\,IV] flux, about half of the 
[Ne\,III] 15.55$\mu$m flux could be due  to a relatively low excitation NLR,
whereas AGN contamination of the stronger [Ne\,II] line is much less
significant. A rough estimate of the intrinsic starburst neon excitation ratio
is hence closer to 0.2 rather than the observed total of 0.39. This is 
in line with the overall picture of an ageing starburst derived on the
basis of near-infrared data, e.g. the low equivalent width of the
near-infrared hydrogen recombination lines (e.g., Tecza et al. \cite{tecza00}).

In the gas-rich circumnuclear region of \object{NGC\,6240}, extinction is
most likely significant. If possible, it should be estimated from 
mid-infrared tracers representative of the starburst activity 
probed by the ISO
observations. The two most reliable estimators, recombination lines and
the [S\,III] lines are of limited use for the \object{NGC\,6240} data, 
however. Because of their low equivalent width, no recombination lines are 
significantly detected in our data.
Comparison of the [S\,III] 18.7$\mu$m line (inside the silicate feature) with
the 33.48$\mu$m line can provide another estimate of obscuration which is 
effectively a lower limit based on the assumption of [S\,III] being in the
low density limit. The 18.71$\mu$m line is indeed undetected, consistent with
high obscuration, but the overall weakness of [S\,III] in \object{NGC\,6240} 
(as in other high-metallicity starbursts, see Verma et al. \cite{verma03}) 
makes the limit on the ratio with [S\,III] 33.48$\mu$m still 
consistent with the
intrinsic ratio, and thus prevents derivation of a lower limit on obscuration.
Another way for obtaining an estimate for the obscuration is to compare the
near-infrared Br$\gamma$ flux of Goldader et al. (\cite{goldader97}),
obtained in a $3\arcsec\times 9\arcsec$ N-S aperture, with
the [Ne\,II] flux. The ratio of 0.012 is a factor of 2 lower than 
the one derived in matching apertures for  Br$\gamma$ and [Ne\,II] fluxes in
M\,82 (F\"orster-Schreiber et al. \cite{foerster01}). This suggests 
10-15A$_{\rm V}$ higher extinction towards NGC\,6240 than the well-studied
extinction towards M\,82 (A$_{\rm V}\approx$5 for the screen case, 
F\"orster-Schreiber et al. \cite{foerster01}), in
total 15-20A$_{\rm V}$ in the simplifying screen assumption. This corresponds
to $\sim$1mag extinction of [Ne\,II] and [Ne\,III]. This estimate is 
very uncertain for several reasons. It is sensitive
to uncertainties in the flux of the weak Br$\gamma$ line that is difficult
to measure accurately on top of the stellar continuum. Adopting
the smaller Br$\gamma$ flux of $1.15\times 10^{-21}$\,W\,cm$^{-2}$ obtained
by integrating over a 5\arcsec\ diameter aperture in the Br$\gamma$ map of 
Tecza et al. (\cite{tecza00}) increases A$_{\rm V}$ by another $\approx$9mag.
The estimate also relies on the assumption of similar metallicities in 
\object{NGC\,6240} and \object{M\,82}. Our obscuration estimate is consistent
with an estimate from the beginning of this century of 
N$_{\rm H}\approx 2\times 10^{22}$ (A$_{\rm V}\approx$10) for the 
atomic hydrogen column 
in front of the radio continuum sources (Beswick et al. \cite{beswick01}). 
Since this column refers only to atomic gas, it is effectively a lower limit
to which any molecular column would have to be added. 

The picture of a low excitation starburst at obscuration somewhat above 
or similar to that of M\,82 is also consistent with the 4$\sigma$ detection of 
the low excitation [N\,II]122$\mu$m line, at a wavelength much less 
sensitive to obscuration. The ratio of 0.14 to [Ne\,II] is slightly above
the estimated ratio 0.12 in M\,82 (Colbert et al. \cite{colbert99} vs. 
aperture-corrected F\"orster-Schreiber et al. \cite{foerster01}). The small
critical density of the [N\,II] line could additionally affect this ratio.
The limits
on the higher excitation lines above 50$\mu$m ([N\,III], [O\,III])
are close to fluxes expected for an M\,82-like excitation; slightly deeper
spectra should detect these lines. 

\subsection{Extremely strong emission of warm molecular hydrogen}
\label{sect:h2}

\begin{figure}
\includegraphics[width=\columnwidth]{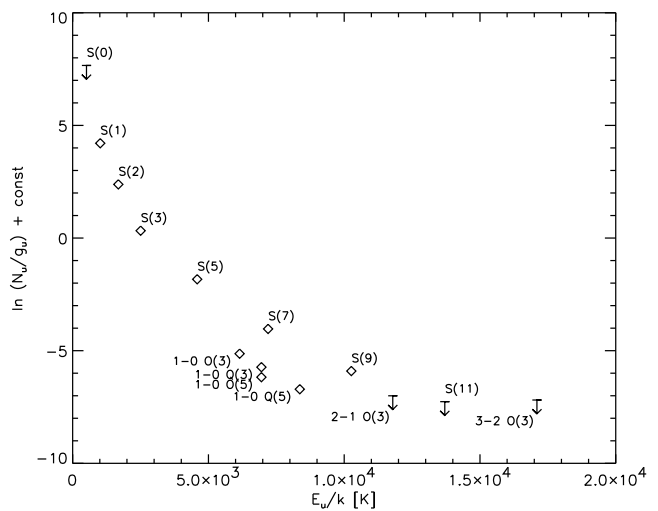}
\caption{H$_2$ excitation diagram for \object{NGC\,6240}, showing for the
lines observed with SWS the upper level population divided by the level 
degeneracy, as a function of upper level energy. The level populations are
derived from observed fluxes without extinction correction.}
\label{fig:h2excit}
\end{figure}

Even more than the near-infrared rovibrational transitions, the mid-infrared
rotational transitions of H$_2$ are extremely strong in \object{NGC\,6240}.
The observed rotational lines in the range 0-0 S(0) to S(11) 
(Table~\ref{tab:fluxes}) carry a total flux of 
$4.5\times 10^{-19}$\,W\,cm$^{-2}$. 
Adding an approximate 1/3 for the unobserved among the pure rotational 
transitions, this corresponds to a luminosity of the H$_2$ rotational lines
of $1.8\times 10^9$\,L$_\odot$ which is 
$\approx$0.3\% of the infrared (8-1000$\mu$m) luminosity 
of the galaxy. A convenient way to
visualize the properties of the H$_2$ emission is an excitation diagram
(Fig.~\ref{fig:h2excit}, adopting transition probabilities of Turner et al.
\cite{turner77}, see also Wolniewicz et al. \cite{wolniewicz98}). 
Thermally excited H$_2$ at a single temperature and high density will
show all transitions on a straight line in such a diagram. The 
concave curvature seen 
in Fig.~\ref{fig:h2excit} indicates a mix of temperatures, similar to 
excitation diagrams for starbursts and AGN observed with ISO 
(Rigopoulou et al. \cite{rigo02}). 
The rotational emission in \object{NGC\,6240}
is, however, not only scaled up but also shows a stronger population
of the higher rotational levels. At the low energy end, both
the limit T$>$145\,K obtained from the limit on the S(1)/S(0) ratio and the
excitation temperature of $\sim$365\,K derived from the S(1) and S(2) lines
are within the ranges derived from those lines in other galaxies 
observed with ISO. Comparing lines originating at higher and lower energies,
however, the flux ratio S(5)/S(1)=1.56 is larger than in any of the starbursts
and AGN presented by Rigopoulou et al. (\cite{rigo02}).

The upper level of the S(3) line which falls near the center of the 
silicate feature appears somewhat underpopulated in Fig.~\ref{fig:h2excit}, 
suggesting significant extinction of A$_{9.6\mu\rm m}\approx$1mag towards
the H$_2$ emitting region. The levels in the first vibrationally excited
state from which the 1-0 lines arise appear underpopulated as well. 
Extinction may contribute here as well,
since the lines from the first vibrationally excited state are at shorter
wavelengths. For most commonly adopted  extinction laws the obscuration at 
2.4--3.2$\mu$m will be similar or less than the extinction at 9.6$\mu$m,
however.
Less than or about one magnitude of extinction is not sufficient to explain
the difference between the population of the vibrational ground state and
first vibrational excited state. Subthermal population of the vibrational
levels at not too high densities may play a role. 

A total mass of 2--$5\times 10^8$\,M$_\odot$ for the `warm' (a few hundred K) 
molecular hydrogen  can be estimated from the observed S(1) flux, and making 
the plausible assumption that the S(1) emission originates in gas that on 
average is 50-100\,K colder than the excitation temperature estimated from
S(2)/S(1). This is a few percent of
the total molecular gas mass estimated from CO observations (Solomon et al. 
\cite{solomon97}, Tacconi et al. \cite{tacconi99}), a fraction not
uncommon in starbursts and AGN. More significantly, 
$\approx5\times 10^6$\,M$_\odot$ of gas at an excitation temperature of
$\sim$1100K are needed to produce the observed S(5) flux.

Draine \& Woods (\cite{draine90}) proposed a model where most of the 
near-infrared rovibrational H$_2$ emission of NGC\,6240 is due to
molecular gas illuminated by shock-produced soft X-rays. Such gas would
also contribute to a lesser degree to the observed rotational line 
emission. An observational
test for this model is the expected emission of H$_3^+$, e.g. at 4.35 and
16.33$\mu$m. We do not detect these lines but the derived 3$\sigma$ limits 
(Table~\ref{tab:fluxes}) are factors of 19 and 4 respectively above the 
predictions. They do not constrain the X-ray illumination scenario, also 
considering the much lower predictions of Maloney et al. (\cite{maloney96}). 
 
Radiative excitation of molecular hydrogen by UV illumination in moderate 
density PDRs and 
subsequent cascading leads to a strong population of the higher vibrationally
excited levels which will be found above the pure rotational lines of similar
energy in an excitation diagram (e.g., Fig.~3 of Timmermann et al. 
\cite{timmer96}; Sternberg \& Neufeld \cite{sternberg99}, Fig.~6 of 
Draine \& Bertoldi \cite{draine99}).
The limits set on the 2-1~O(3) and 3-2~O(3) lines rule out a dominant
contribution from such moderate density PDRs for the \object{NGC\,6240}
H$_2$ spectrum. The higher vibrational levels are clearly not overpopulated
with respect to the lower ones. This agrees with conclusions that have
been reached on the basis of near-infrared spectra (e.g. van der Werf et al.
\cite{vdwerf93}, Sugai et al. \cite{sugai97}). 

Beyond excluding UV excitation in low density PDRs as the prime contributor, 
the thermal H$_2$ spectrum observed in the SWS range holds little direct 
clue as to its sources of excitation. However, the detailed multiwavelength 
observations available for \object{NGC\,6240} make a strong contribution
of shocks likely. The CO velocity field
in the inner gas concentration is highly turbulent, suggesting
rapid dissipation in this large gas mass (Tacconi et al. \cite{tacconi99}).
Shocks that will occur under such conditions are efficient 
sources of rotational H$_2$ emission up to fairly high excitations, with the
excitation diagram generally showing a smooth trend with level energy 
(e.g., OMC-1, Figs. 2 and 5 of Rosenthal et al. \cite{rosenthal00}). 
The [O\,I] 63$\mu$m
line expected to be strong in shocks (Hollenbach \& McKee \cite{hollenbach89})
is also observed to be prominent in
\object{NGC\,6240} (Table~\ref{tab:fluxes}). Its flux is $\sim$2.5 times that
of [C\,II] 157$\mu$m, unusually high compared to other starbursts and
IR-bright galaxies (e.g., Fischer et al. \cite{fischer99}, Colbert et al.
\cite{colbert99}). This high [O\,I]/[C\,II] ratio is even above 
the relatively  high [O\,I]/[C\,II] ratios observed in some [C\,II] 
deficient galaxies
(Malhotra et al. \cite{malhotra01}, Luhman et al. \cite{luhman03}) which
are often ascribed to intensely heated (high G$_0$/n) PDRs. The very high 
ratio for \object{NGC\,6240} is probably not more than partially of 
the same origin - at $log(L_{\rm [CII]}/L_{\rm FIR})$=-2.7 
(see also Stacey et al. \cite{stacey91},
Luhman et al. \cite{luhman98}) NGC\,6240 is not [C\,II]-deficient.
Morphology, kinematics and excitation conditions of the near-infrared
rovibrational H$_2$ lines have been variously used to argue
for shock excitation of these lines as well (e.g., van der Werf 
\cite{vdwerf93}; Sugai et al. \cite{sugai97}). Single velocity/density 
C shock models (Kaufman \&
Neufeld \cite{kaufman96}) may not be able to reproduce simultaneously
all the rotational lines. This can be plausibly accounted for by a mix of 
speeds and by a contribution of PDRs to the lowest rotational transitions.

Depending on shock conditions,  cooling by the pure rotational lines of 
H$_2$ will correspond to a few percent to tens of percent of the mechanical
luminosity of the shock (e.g. Kaufman \& Neufeld \cite{kaufman96}). The 
mechanical luminosity corresponding to the observed rotational H$_2$ cooling
may thus be approaching 10$^{10}$L$_\odot$. A refined estimate is possible
considering other potential major shock cooling lines. We do not detect in the
LWS spectrum lines of CO and H$_2$O that can also be major coolants for 
shocks in molecular regions. Table~\ref{tab:fluxes} includes limits for CO 
and H$_2$O lines in the most sensitive regions of the spectrum. These are 
factors $\sim$5 less than the fluxes of the strongest rotational 
molecular hydrogen lines. We may use these limits and the ratio S(5) to
1-0 Q(3) to search for acceptable conditions in the range of shock speeds
and densities sampled by the shock models of Kaufman \& Neufeld 
(\cite{kaufman96}). These argue for, on average, relatively fast
(V$\sim$30\,km/s) shocks in gas of preshock density $\sim$10$^5$\,cm$^{-3}$.
Under these conditions, the rotational lines of molecular hydrogen will
radiate $\sim$40\% of the shock mechanical luminosity. 

The strong and probably shock dominated [O\,I]63$\mu$m line indicates 
that this picture is not complete. It radiates another 0.3\% of the 
bolometric luminosity but is not produced in significant quantities in the
nondissociative (C-) shock models cited above, where oxygen is assumed mostly 
locked in molecules (but cf. Poglitsch et al. \cite{poglitsch96}). 
Strong [O\,I] emission is however a key result of models of dissociative
(J-) shocks (Hollenbach \& McKee \cite{hollenbach89}). A single shock model
is not adequate to reproduce a galaxy with a wide range of shock speeds and
preshock densities. Assuming that the H$_2$ rotational lines and [O\,I]
are the strongest coolants of shocks in \object{NGC\,6240}, the total 
mechanical luminosity of these shocks must be of the order 
10$^{10}$\,L$_\odot$ even if some fraction of those lines has other origins.

If derived from the
kinetic energy of the turbulent central gas concentration 
($\sim 3\times 10^9$\,M$_\odot$, $\sigma\approx$150\,km/s, Tacconi et al. 
\cite{tacconi99}), a shock mechanical luminosity of 10$^{10}$\,L$_\odot$
will dissipate this turbulent energy within a few
million years, i.e. on the dynamical timescale of the central gas
concentration. Additional contributions may come from the mechanical
luminosity of the starburst `superwind' in \object{NGC\,6240}, estimated
to be of the order 10$^{10}$L$_\odot$ (e.g., Heckman et al. 
\cite{heckman90}, Schulz et al. \cite{schulz98}).

\section{What are the contributions of star formation and AGN to
the luminosity of \object{NGC\,6240}?}

We discuss several approaches
to quantify the luminosity of the starburst and the AGN(s) in 
\object{NGC\,6240}. All these approaches directly relate a mid-infrared or
X-ray observable to the infrared (or bolometric) luminosity. In this 
they differ
from the purely mid-infrared diagnostic diagrams (Genzel et al.
\cite{genzel98}; Laurent et al. \cite{laurent00}) where extrapolation
from mid-infrared to bolometric has to be considered separately (e.g.
Sect.~3.6 of Genzel et al. \cite{genzel98}). The uncertainties
will thus include the uncertainties of the extrapolation to bolometric 
luminosity.
In addition to normal uncertainties in measurements and methods, a consistent
difficulty of several of these approaches is incomplete
characterisation of obscuration. Effectively, this leads to a situation
where a lower limit to the luminosity of a component can be set with
some confidence, but where it is very difficult to assess whether
more of this component is present at higher obscuration or whether a 
different component takes over.
It is then more useful to quantify what is observed, effectively defining
a lower limit, than to make possibly biased interpretations of the 
obscured components.
It is also worth remembering that obscuration
towards spatially distinct regions (e.g., starburst, Narrow Line Region, 
AGN dust source, AGN X-ray source) is likely to differ considerably.

Since modelling from first principles is currently unreliable for several of
the constraints discussed below, we follow an approach of quantitative but 
empirical comparison to starburst and AGN templates. For \object{NGC\,6240} 
as well as for the (dusty) comparison starbursts we assume the bolometric
luminosity to be equal to the 8-1000$\mu$m infrared luminosity, while the
relation between these quantities is explicitly included for AGN.
For brevity, we do not
include the radio emission of \object{NGC\,6240} in our discussion. The 
radio results summarized by Beswick et al. (\cite{beswick01}) are 
consistent with coexistence of starburst and AGN, as outlined below.

\subsection{Star formation: Low excitation fine-structure lines}
With [Ne\,II] mainly excited by star formation, this line can be used 
to estimate
the ionizing luminosity and bolometric luminosity of the starburst.
During the evolution of a starburst, the ratio of ionizing luminosity
of the hot stars to bolometric luminosity will decay, as will the
excitation observed in the mid-infrared fine-structure lines (see, e.g.,
the models presented in Thornley et al. \cite{thornley00}). When trying 
to estimate
the bolometric luminosity of the starburst component from
the observed low excitation fine structure lines, it is hence 
important to choose templates with excitation similar to the ratio 
[Ne\,III] 15.55$\mu$m / [Ne\,II] 12.81$\mu$m $\approx$0.2 derived above
for the \object{NGC\,6240} starburst. Of the starbursts observed
by Verma et al. (\cite{verma03}) this ratio is between 0.1 and 0.4 for
\object{M\,82}, \object{NGC\,3256}, and \object{NGC\,3690}/\object{IC\,694}. 
For these
galaxies, we compute a mean $log(L_{\rm [NeII]}/L_{\rm IR})=-3.03$. The values
for the individual galaxies are between -2.89 and -3.22. To meaningfully 
compare with the IRAS FSC-based infrared luminosities, we
have considered their spatial extent compared to the SWS aperture by 
adding the two apertures for the two pointings on 
\object{NGC\,3690}/\object{IC\,694}
which together fully cover the active regions of this interacting
system (see Fig.~1 in Verma et al. \cite{verma03}) 
and by doubling the
observed flux for \object{M\,82} where the SWS aperture only partially
covers the starburst region (see F\"orster Schreiber et al. \cite{foerster01}).
 
Using this empirical calibration, the observed [Ne\,II] luminosity of 
$4.8\times 10^8$\,L$_\odot$ converts into a starburst bolometric luminosity of 
$5\times  10^{11}$\,L$_\odot$, i.e. the major part of the bolometric 
luminosity of \object{NGC\,6240}.
This would be an underestimate if the obscuration were higher than in the
starburst templates. While there is evidence for higher obscuration
of the \object{NGC\,6240} starburst than in \object{M\,82} 
(Sect.~\ref{sect:sbprop}), we stick with the conservative
assumption of similar obscuration because of the uncertainty of this estimate
and of obscuration estimates for the other templates
\object{NGC\,3256} and \object{NGC\,3690}. The starburst luminosity
would be overestimated if the metallicity were much higher than
for the templates, which are already at supersolar metallicities,
however (Verma et al. \cite{verma03}). It would also be overestimated
if the AGN contribution to [Ne\,III] were less, making the starburst more
highly excited with a correspondingly lower ratio of total to ionizing
luminosity.

\subsection{Star formation: Aromatic emission features}

\begin{figure}
\includegraphics[width=\columnwidth]{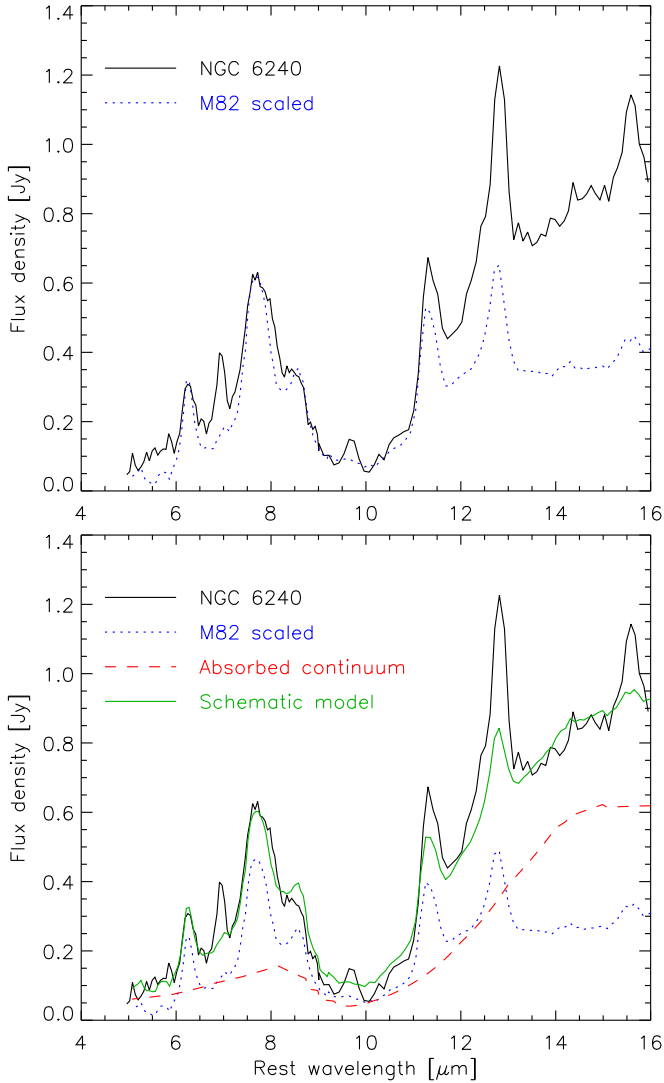}
\caption{Low resolution ISOCAM 5--16$\mu$m spectrum of \object{NGC\,6240} 
compared to
a scaled spectrum of \object{M\,82} (top) and a schematic model consisting 
of the sum of a scaled \object{M\,82} spectrum and an absorbed power law
continuum (bottom). 
Obscuration of the continuum is needed for a satisfactory representation 
including the region of silicate feature. The schematic model is inadequate
in the regions of strong emission lines because of differences in the 
emission line spectra of \object{NGC\,6240} and \object{M\,82}.}
\label{fig:cvfdecomp}
\end{figure}

The mid-infrared aromatic `PAH' emission features are observed over a wide 
range of interstellar medium conditions from the diffuse ISM to star forming
regions but are not observed to be strong in the immediate vicinity of AGN. 
A number of studies have successfully decomposed low resolution
mid-infrared spectra of galaxies into three components: A `PDR' component
dominated by the aromatic emission features, an `H\,II' very small grain
continuum steeply
rising towards longer wavelengths and a hotter `AGN' dust 
continuum (Laurent et al. \cite{laurent00}, see also 
Sturm et al. \cite{sturm00}, Tran et al.
\cite{tran01},  and F\"orster Schreiber et al. \cite{foerster03}).
Dust obscuration can be important (Lutz et al. \cite{lutz98a}, Tran et al.
\cite{tran01}), with the additional complication of ice features
(Spoon et al. \cite{spoon02}). The \object{NGC\,6240} spectrum 
(Fig.~\ref{fig:cvfdecomp}) is
similar to starburst spectra which are a superposition of `PDR' and `H\,II',
but there is an additional continuum extending down to 5$\mu$m, indicative
of an AGN continuum which we will discuss in Sect.~4.4. Subtracting such a
continuum, the peak flux density in the 7.7$\mu$m aromatic feature
is $\approx$0.45Jy.

The physics and chemistry of the carriers of
the aromatic features and their transient heating by UV photons originating
over a very wide range of stellar types is complex and not fully
understood. This certainly provides for a less direct link from PAH 
luminosity to the bolometric luminosity of a starburst than from luminosity of
the emission lines originating in the medium ionized by OB stars to 
bolometric luminosity of the starburst. Nevertheless,
it is worth trying to extrapolate from the PAH emission to the
associated bolometric emission by using starbursts as templates.
In order to minimize aperture effects in the comparison to the IRAS FSC-based
infrared luminosity, we use objects for which PAH data are available covering 
a major part of the IRAS beam and/or the starburst 
region\footnote{ISOCAM-CVF M\,82, M\,83, NGC\,253, NGC\,520, 
NGC\,1808, NGC3690, NGC4038/39, Arp236, NGC\,7252 (O. Laurent. priv. comm.) 
and NGC\,3256 ISOPHOT-S (Rigopoulou et al. \cite{rigo99})}. 
Comparing the PAH 7.7$\mu$m feature peak flux density in Jy, and the 
8-1000$\mu$m IR
flux in W\,m$^{-2}$, we obtain a mean log(S$_{7.7}$/F$_{\rm IR}$) of 11.84 with
a dispersion of 0.25. Application of this relation to \object{NGC\,6240}
results in a starburst luminosity of $2\times 10^{11}$L$_\odot$.  
  
This is an underestimate if the PAH emission is more obscured than
in the comparison starbursts. Indications for the corresponding 
relative weakening of the 8.6 and 11.3$\mu$m features (in the wings of 
the silicate absorption) are not strong, however (Fig.~\ref{fig:cvfdecomp}).
More important for assessing the robustness of this estimate are  
variations of the aromatic feature emission with environment. It is known
since the IRAS mission that the aromatic features decrease when 
approaching a hot star or when
going inwards from PDRs to H\,II regions (e.g., Boulanger et al. 
\cite{boulanger88}; Verstraete et al. \cite{verstraete96}). Correspondingly,
trends are observed in `normal star forming' galaxies between the average
radiation field intensity, echoed in the 60/100$\mu$m flux ratio  tracing 
the large grain temperature, and the ratio of PAH emission to total infrared 
luminosity (Dale et al. \cite{dale01}). This introduces additional 
uncertainties when extrapolating from PAH features to total luminosity and
may cause underestimates if much of the star formation is concentrated 
in a small region.

\subsection{AGN: High excitation lines from the Narrow Line Region}
Here, we use the Narrow Line Region [O\,IV] 25.89$\mu$m luminosity as an 
indicator of the AGN luminosity and use again an empirical calibration.
To avoid biasing our result by star formation contributions to the
bolometric luminosity of comparison objects, we choose as comparison those 
among the Seyferts observed by Sturm et al. (\cite{sturm02}) that have
measured [O\,IV] lines and are not listed in their Table~1 as having strong
evidence for circumnuclear star formation. The mean 
log(L$_{\rm [OIV]}$/L$_{\rm IR}$) of these 16 objects is -3.01, with 
a dispersion of
0.29. Using this calibration, the observed [O\,IV] luminosity of 
$7\times 10^7$\,L$_\odot$ converts to an AGN infrared luminosity of 
$7\times 10^{10}$\,L$_\odot$. This is an extrapolation based on
the relation of [O\,IV] and {\em infrared} luminosities in the comparison
Seyferts. In Seyfert galaxies, the 10-1000$\mu$m infrared range will represent
very roughly 1/2 of the bolometric luminosity (Spinoglio et al.
\cite{spinoglio95}), with significant emission emerging at shorter wavelengths
in the near-infrared, visual, and UV to X-ray ranges. The shorter 
wavelength emission in the integrated SEDs of Seyferts as in Spinoglio et al.
(\cite{spinoglio95}) will not always be pure AGN emission because 
of host contributions, but we assume it is dominated by the AGN. 
For an obscured AGN in \object{NGC\,6240}, it is
then plausible to assume that much of the NIR/optical/UV/soft-X
part of the AGN bolometric luminosity would be absorbed and reradiated in 
the mid- and far-infrared.
We hence double the estimate of the AGN contribution to the infrared  
luminosity to $1.4\times 10^{11}$\,L$_\odot$.

This is an overestimate if the non-AGN shock contribution to [O\,IV] were 
large in \object{NGC\,6240}, contrary to the estimate in 
Sect.~\ref{sect:nlrdet}.
While we have tried to exclude objects with PAHs indicating strong star 
formation from the comparison sample, it may still be an overestimate if 
the IR emission of the comparison objects still has significant 
contributions from star formation. This is possible because of
aperture effects between the PAH measurements and the FIR continuum
measurements in currently available data. An example for this problem is 
NGC\,1068 with very weak PAH in the
ISOPHOT beam, and considerable star formation outside this beam
but inside the IRAS beam (Telesco \& Decher \cite{telesco88}, Le Floc'h et al. 
\cite{lefloch01}). Similar effects are hard
to firmly exclude for other less well studied large galaxies because of the 
mismatch between the IRAS beam and the ISOPHOT-S aperture used to search 
for PAHs indicating star formation. High spatial resolution both in
mid-infrared spectrophotometry and in the far-infrared would be needed
to minimize these problems.
Our AGN luminosity is an underestimate if the NLR of \object{NGC\,6240} is 
significantly 
obscured even at 26$\mu$m. Several magnitudes of visual extinction needed
to hide the NLR in the optical do not yet have a significant effect, however.

\subsection{AGN: Mid-Infrared continuum}

Mid-infrared spectra of AGN show a strong continuum due to warm dust
in the vicinity of the AGN, either in the putative torus or  on larger
scales e.g. embedded in the Narrow Line Region. The spectrum of 
\object{NGC\,6240} has excess continuum over a pure starburst spectrum
which may either be AGN related or an increased contribution of the steep
`H\,II' continuum. The excess continuum must be obscured in order not to
violate the observed spectrum in the region of the silicate feature.
Because of this obscuration and the steep rise of a starburst-related
`H\,II' excess, an excess AGN continuum is best quantified shortwards
of the PAH complex, at $\sim$5.9$\mu$m (Laurent et al. \cite{laurent00}). 
Subtracting from the observed
spectrum of \object{NGC\,6240} a scaled M\,82 spectrum reproducing 
the 6.2 and 7.7$\mu$m PAH
features, we estimate the excess AGN continuum to be $\sim$0.07Jy at 5.9$\mu$m
(Fig.~\ref{fig:cvfdecomp}).

Again we use AGN without evidence for circumnuclear star formation (no or only
weak PAH features according to ISOPHOT-S observations) as 
templates. From 40 such AGN with z$<$0.1 and observations in the ISO archive we
estimate a mean relation between 5.9$\mu$m continuum flux density in Jy, and 
the IR flux in W\,m$^{-2}$, of
 log(S$_{5.9}$/F$_{\rm IR}$)=11.70 with
a dispersion of 0.29. Application of this relation to \object{NGC\,6240}
results in an AGN infrared luminosity of $4.1\times 10^{10}$L$_\odot$. 
Again, we have to consider that for the comparison Seyferts a noticeable
part of the bolometric will emerge at short wavelengths that are likely 
absorbed and redistributed to the infrared in \object{NGC\,6240}. As 
for the estimate based on the [O\,IV] line, we apply an additional
factor 2 correction to account for the difference between infrared and
bolometric luminosity in the comparison Seyferts, arriving at an AGN
bolometric luminosity of $\approx8\times 10^{10}$L$_\odot$.
  
This is an underestimate if the AGN dust continuum in \object{NGC\,6240} is 
heavily obscured. 
Relatively little obscuration is sufficient to produce the silicate absorption
needed for a spectrum fit (e.g. A$_{\rm V}\approx$20 in 
Fig.~\ref{fig:cvfdecomp}), but high values are not excluded. Another 
potential bias which can cause an underestimate of the AGN luminosity 
is due to the choice of PAH-free comparison
objects. If unification related orientation effects cause variations
in the AGN mid-IR brightness (Clavel et al. \cite{clavel00}), then our 
comparison objects selected to have a bright continuum standing out of 
the host galaxy PAHs might 
be above average in mid-IR brightness compared to an orientation-unbiased 
sample. In the absence of observations that can spatially separate host and
AGN at mid {\em and} far-infrared wavelengths this bias is hard to break.
Again, we will have overestimated the AGN luminosity if the IR luminosities
of the comparison objects still contain significant starburst contributions,
outside the ISOPHOT-S beam but inside the IRAS beam. We will also have
overestimated the AGN luminosity if some of the excess 5.9$\mu$m continuum
is still H\,II region related. 

\subsection{AGN: X-ray emission}
The X-ray emission of \object{NGC\,6240} holds important clues to the nature
of its AGN. The soft X-ray emission (0.5-2\,keV) is still dominated by 
extended thermal
emission tracing the starburst superwind (Komossa et al. \cite{komossa98};
Iwasawa \& Comastri \cite{iwasawa98}). The hard X-rays (2-10\,keV) are
dominated by AGN emission reflected by cold and warm matter
(Iwasawa \& Comastri 
\cite{iwasawa98}) from two AGN located in the two nuclei (Komossa et al. 
\cite{komossa03}). Above 10\,keV, absorbed direct AGN emission has 
been observed
with BeppoSAX and RXTE (Vignati et al. \cite{vignati99}; Ikebe et al.
\cite{ikebe00}). The question is to
estimate the bolometric luminosity of the AGN from the observed direct
and/or reflected AGN X-rays. 

Spectral fits to the complete X-ray spectrum of \object{NGC\,6240} suggest 
an intrinsic (corrected for absorption) 2-10\,keV luminosity of 
$2.2\times 10^{43}$ to $2.7\times 10^{44}$\,erg\,s$^{-1}$ 
(Vignati et al. \cite{vignati99}, Ikebe et al.
\cite{ikebe00}; corrected to the distance assumed in this paper). The main 
uncertainties involved in the analysis of the 2-10\,keV emission are the 
geometry of the 
X-ray reflector, and the potential effect of scattering by the absorbing
matter. Similar uncertainties arise when extrapolating from the directly 
observed very hard ($>$10\,keV) X-ray emission, due to the uncertain 
spectral index towards softer X-rays.
In a next step, the 2-10\,keV luminosity has to be extrapolated to the
bolometric luminosity. Assuming the mean radio-quiet quasar SED of 
Elvis et al. (\cite{elvis94}, their Fig.~10), 
L$_{2-10\rm keV}$/L$_{\rm Bol}\approx$0.09,
with significant scatter around this mean relation. In the regime of 
Seyferts,  L$_{2-10\rm keV}$/L$_{\rm IR}$ is typically around 0.1 for a major 
fraction of objects that are not absorbed in the 2-10keV range, with a tail to 
smaller ratios (Risaliti et al. \cite{risaliti00}). We have verified
this by comparing observed 2-10keV fluxes of z$<$0.1
Seyfert 1s from George et al. (\cite{george98}) and extinction-corrected 
2-10keV fluxes of Seyfert 2s from Bassani et al. (\cite{bassani99}) with 
infrared fluxes. We have
restricted the samples to objects where the absence of PAHs in ISOPHOT 
spectra suggests small star formation contribution, and excluded Compton
thick objects. The mean  log(L$_{2-10\rm keV}$/L$_{\rm IR}$) for 18 objects 
(12 Sy1, 6 Sy2) is -1.0 with a dispersion of 0.29. Following our previous 
approach, we assume $L_{\rm IR}\approx 0.5 L_{\rm Bol}$ for the Seyferts.

In total, the AGN luminosity estimated from the X-ray spectroscopy is
in the range $2\times 10^{44}$ to $6\times 10^{45}$\,erg\,s$^{-1}$, or 
between 10\% and several times the luminosity of \object{NGC\,6240},
depending on the estimate of the X-ray luminosity and assumptions for the
bolometric correction. The X-ray data
suggest the AGN plays a significant and possibly dominant role but cannot
pin down the AGN luminosity accurately.

\subsection{Combining the constraints}

\begin{table}
\begin{tabular}{lcc}\hline
Constraint                  &Starburst    &AGN      \\ \hline
Low excitation mid-IR lines &50--100\%    &         \\
PAH features                &$>$33\%      &         \\
High excitation mid-IR lines&             &$>$24\%  \\
Mid-IR dust continuum       &             &$>$13\%  \\
Hard X-ray emission         &             &10--100\%\\ \hline
\end{tabular}
\caption{Constraints on the contribution of starburst and AGN activity to
the infrared luminosity of \object{NGC\,6240}.}
\label{tab:constraints}
\end{table}

The five constraints on the contributions of star formation and AGN in
\object{NGC\,6240} discussed in the previous subsections are summarized in 
Table~\ref{tab:constraints}. We assign highest weight to two constraints. 
The X-ray constraint suggests that the AGN is contributing
a significant and possible dominant fraction of the luminosity, the latter
situation being within the uncertainty of the estimate based on the 
{\em observed}
X-ray emission rather than just being allowed as an unseen obscured component.
The second key constraint is the one on the starburst
luminosity derived from the low excitation fine-structure lines, which is based
on relatively straightforward analysis of starburst H\,II regions. 
It is difficult to envisage H\,II regions that are low excitation and still 
have a high ratio of ionizing luminosity to total luminosity, which would
be needed to make the starburst contribution to the bolometric luminosity
insignificant. We 
suggest that the most likely range of contributions to the luminosity of
\object{NGC\,6240} is 50 to 75\% starburst and 25 to 50\% AGN. This estimate
is based on direct comparison of mid-IR and X-ray observables to the 
bolometric 
luminosity. Reassuringly, the agreement with pure mid-IR diagnostics
(Genzel et al. \cite{genzel98}; Laurent et al. \cite{laurent00}) is good.

\section{Conclusions}
Infrared spectroscopy of \object{NGC\,6240} detects signatures of powerful
star formation activity and of an obscured AGN. We have used these
results in conjunction with X-ray spectroscopy to estimate the 
contributions to the total luminosity of starburst (50 to 75\%) and AGN 
activity (25 to 50\%). 
\object{NGC\,6240} is a luminous source of rotational
emission of molecular hydrogen amounting to 0.3\% of the bolometric luminosity,
a similar amount is radiated in the [O\,I] 63$\mu$m line.
Shocks due to the turbulent central velocity field and the superwind are
likely to dominate these extraordinary levels of emission, and imply 
a mechanical luminosity of at least 10$^{10}$\,L$_\odot$.

\begin{acknowledgements}
    We acknowledge support by Verbundforschung (50 OR 9913 7), for the
    ISO spectrometer data center at MPE by DLR (50 QI 0202), and by
    the German -- Israeli Foundation (grant I-0551-186.07/97). We thank
    Olivier Laurent for access to his ISOCAM-CVF spectra,  
    Matthias Tecza for providing Br$\gamma$ data, and Stefanie Komossa
    for discussions. We are grateful to the referee for valuable suggestions.
\end{acknowledgements}

\end{document}